\documentclass[referee,letterpaper]{raa}           
\usepackage{graphicx,times}
\usepackage{natbib}
\usepackage{amssymb,amsmath}
\usepackage{pdfpages}

\def\myfigwidthS{0.75\textwidth}  
\def\myfigwidthL{0.95\textwidth}  

\begin{document}

   \title{The inner solar system cratering record and the evolution of impactor populations}

 \volnopage{ {\bf 2012} Vol.\ {\bf X} No. {\bf XX}, 000--000}
   \setcounter{page}{1}

   \author{Robert G. Strom\inst{1}, Renu Malhotra\inst{1}, Zhiyong Xiao\inst{2}, Takashi Ito\inst{3}, Fumi Yoshida\inst{3}, Lillian R. Ostrach\inst{4}
   }

   \institute{Lunar and Planetary Laboratory, The University of Arizona, Tucson, Arizona 85721, USA; {\it rstrom@lpl.arizona.edu}\\
        \and
             Faculty of Earth Sciences, China University of Geosciences (Wuhan), Wuhan, Hubei, China \\
        \and
             National Astronomical Observatory, Osawa 2--21--1, Mitaka, Tokyo 181--8588, Japan \\
        \and
             School of Earth and Space Exploration, Arizona State University, Tempe, Arizona, USA \\
\vs \no
   {\small Received 20XX XXXX XX; accepted 20XX XXXX XX}  
}

\clearpage

\abstract{
We review previously published and newly obtained crater size-frequency distributions in the inner solar system.   These data indicate that the Moon and the terrestrial planets have been bombarded by two populations of objects. Population 1, dominating at early times, had nearly the same size distribution as the present-day asteroid belt, and produced the heavily cratered surfaces with a complex, multi-sloped crater size-frequency distribution. Population 2, dominating since about 3.8-3.7 Ga, has the same size distribution as near-Earth objects (NEOs), had a much lower impact flux, and produced a crater size distribution characterized by a differential -3 single-slope power law in the crater diameter range 0.02 km to 100 km. Taken together with the results from a large body of work on age-dating of lunar and meteorite samples and theoretical work in solar system dynamics, a plausible interpretation of these data is as follows.  The NEO population is the source of Population 2 and it has been in near-steady state over the past $\sim$~3.7-3.8 gigayears; these objects are derived from the main asteroid belt by size-dependent non-gravitational effects that favor the ejection of smaller asteroids.  However, Population 1 were main belt asteroids ejected from their source region in a size-independent manner, possibly by means of gravitational resonance sweeping during giant planet orbit migration; this caused the so-called Late Heavy Bombardment (LHB).  The LHB began some time before $\sim$~3.9 Ga, peaked and declined rapidly over the next $\sim$~100 to 300 megayears, and possibly more slowly from about 3.8--3.7 Ga to $\sim$~2 Ga.  A third crater population (Population S) consists of secondary impact craters that can dominate the cratering record at small diameters. 
}

   \authorrunning{R. G. Strom et al. }            
   \titlerunning{The inner solar system cratering record}  
   \maketitle

\section{Introduction}
\label{sec:intro}

It is often implied that there is only one crater population
in the Solar system caused by one population of impacting
objects. Some authors
\citep[e.g.][]{Hartmann:1995,Hartmann:2001,Ivanov:2002} maintain that the
crater size-frequency distribution (SFD) characteristic of the ancient period of
heavy bombardment has persisted throughout Solar system history.
The commonly used scheme for determining absolute model ages
derived from the impact cratering record is based on this assumption
\citep[e.g.][]{Neukum:2001,Hartmann:2005,Michael:2013}.

\begin{figure} \centering
\includegraphics[width=\myfigwidthL, angle=0]{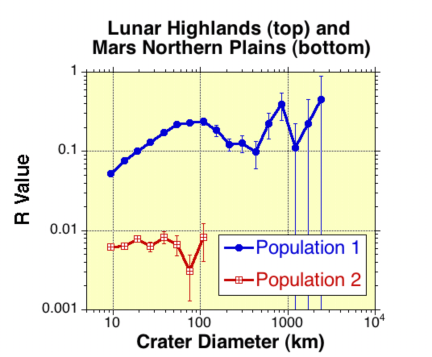}
\caption[]{%
This $R$ plot shows the complex SFD characteristic of Population 1
craters, represented here by the lunar highlands craters, and the nearly horizontal
curve (equivalently, a simple $-3$ power law of the SFD)
of Population 2 craters represented here by the craters on the Northern Plains of Mars.
}
\label{fig:fig01}
\end{figure}

Based on crater SFDs, however, \cite{Strom:2005} 
suggested that the inner Solar system planets and the Moon have been 
bombarded by two different populations of objects, distinguishable by their
size-frequency distributions. 
Following \citet{Strom:2005}, we refer to these as Population 1 and
Population 2. Population 1 is associated with the heavily cratered surfaces
on Mercury, the Moon, and Mars; it has a complex size distribution of
craters, characterized  by a differential $-2.2$ slope at diameters less
than about 50 km, a nearly flat part (differential slope $-3$) between
50 and 100 km, and sloping downward to the right (differential slope
$-4$) at diameters between about 100 km to 300 km. It also appears to
slope
 upward at diameters between about 400 and 1000 km.
Population 2 is associated primarily with
lightly cratered, younger plains units on the Moon, Venus, Mars, and
Mercury; these craters are characterized by a single-slope
differential $-3$ power-law size distribution which plots as a nearly
straight horizontal line on an $R$ plot. The crater density of
Population 1 exceeds that of Population 2 by more than an order of
magnitude for crater diameters greater than 10 km.
Figure \ref{fig:fig01} shows the Relative plots (``$R$ plots'', 
see section \ref{sec:Rplot}] of examples of these
two crater populations. 

The much higher density of Population 1 indicates that it is a
much older impact record than Population 2. The differing shapes of
the two populations suggest the possibility of different dynamical
origins of the Population 1 (old) and Population 2 (younger) impactor
populations. The absence of Population 1 on younger terrains implies
that the SFD of inner solar system impactors changed
from Population 1 to Population 2 at some ancient time. 

Distinguishing between the two crater populations and understanding
the time evolution of the impactor populations is important to
understanding the dynamical history of the Solar system. It is also
directly related to the reliability of the prevalent age dating
technique using crater counts \citep{Michael-Neukum:2010}. 
Many of the crater SFDs published in 
various journals and book chapters over the past $\sim40$ years were
presented before it was realized that two crater populations were
present on the Moon and the terrestrial planets, and that they
differed significantly from crater populations on the outer planet
satellites. During most of this time the origin of the objects
responsible for the cratering record was not well understood. Also, it
was not clearly understood how widespread and important were secondary
impact craters to the small crater population, at diameters $D\lesssim 1$ km on the Moon and Mars and $\lesssim10$ km on Mercury.  

In the present paper we review and provide an updated and
detailed record of the data on the two crater populations and an 
updated discussion of their implications for the time evolution of 
impactor populations in the inner solar system; these were presented
only in abbreviated summary form in \cite{Strom:2005}.
We also present newly obtained MESSENGER data for craters on 
Mercury, new crater counts on the Orientale basin on the 
Moon and new counts of small rayed craters on Mars.  These new data 
and an analysis of secondary impact craters augment and support
the previously published crater SFDs found on the 
inner solar system planets.  

This paper is organized as follows.  In Section 2 we describe in some
detail the definition and usefulness of the $R$ plot.  In Section 3, we give
a comprehensive summary of the crater SFDs
found on each of the terrestrial planets and the Moon, as well a discussion of the
secondary crater data; as a foil to the inner solar system crater record, we also provide (in section 3.4) a summary of the crater SFDs found on outer planet satellites.  In Section 4, we
discuss the implications of the crater record and its interpretation
in theoretical models of the dynamical history of the solar system,
including the putative spike in the impact flux known as the `Late Heavy Bombardment' (LHB)
that is thought to have occurred at about $\sim3.9$ Ga~\citep{Turner:1973,Tera:1974}. 
A list of references and notes for our sources of data as well as two supplementary figures are provided in the Appendix.

\section{The Relative Size-Frequency Distribution Plot}
\label{sec:Rplot}

\begin{figure} \centering
\includegraphics[width=\myfigwidthL, angle=0]{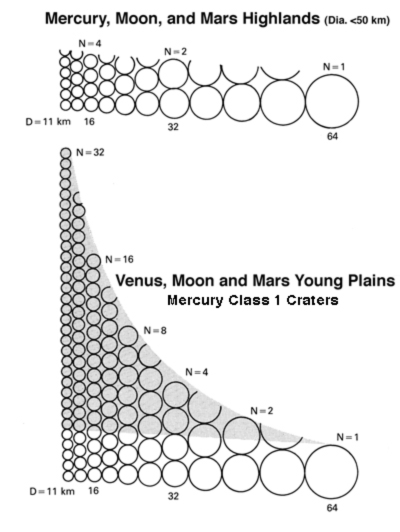}
\caption[]{%
This diagram graphically shows the difference between a differential
$-2$ power law (top) that is characteristic of Population 1 at diameters
less than about 50 km and a differential $-3$ power law characteristic of
Population 2 (bottom). Each circle represents a crater of a given
diameter. The shaded area in the bottom diagram is the difference
between the two populations.
}
\label{fig:fig02}
\end{figure}

The ``Relative'' plot (or $R$ plot) method of displaying the crater and
projectile SFDs is used throughout this
paper. The $R$ plot was devised by the Crater Analysis Techniques
Working Group \citep{Arvidson:1979} to better show the size
distribution of craters and crater number densities for determining
relative ages. When sufficiently large and accurate data sets are
available, the $R$ plot provides a more critical and sensitive
comparison between SFDs than cumulative
plots. The latter tend to smear out important details of the crater
SFD curves and can lead to erroneous interpretations. 
This happened frequently in the 1960s
and 1970s, which led to the formation of a NASA Working Group to
remedy the problem. Figure \ref{fig:fig02} is a diagrammatic
representation of a differential $-2$ and $-3$ size distribution for
diameters between 11 and 64 km diameter.
This figure illustrates visually the large difference between
differential $-2$ and $-3$ power law distributions. 
The significant differences between these distributions were sometimes overlooked owing to the use of cumulative plots (e.g., \citet{Hartmann:1966,Wilhelms:1978,Michael:2013}).
The $R$ plot was strongly recommended by the Crater Analysis Techniques
Working Group \citep{Arvidson:1979}, in addition to any other plots
authors chose to use.  

On an $R$ plot, the SFD is normalized to a power
law differential size distribution function, $dN(D) \sim D^p dD$,
where $D$ is diameter.  The index $p=-3$ is recommended because most crater SFDs are observed to be piecewise within $\pm 1$ of a $p = -3$ power law
distribution. 
The discretized equation for the $R$ value is: $R = D^3 N/A(b_2 - b_1)$;
where $D$ is the geometric mean diameter of the size bin $\sqrt{b_1 b_2})$,
$N$ is the number of craters in the size bin, $A$ is the area over which the
counts were made, $b_1$ is the lower limit of the size bin, and $b_2$ is the
upper limit of the size bin. Usually, although not required, the size
bins are in $\sqrt{2}$ increments because there are many more small craters
than large craters. 
In an $R$ plot, $\log_{10} R$ is plotted on the $y$-axis and $\log_{10} D$ is
plotted on the $x$-axis. Thus, a $p = -3$ SFD plots as a
horizontal straight line; a $p = -2$ SFD slopes down to the
left at an angle of $45^\circ$, and a $p = -4$ SFD slopes down to the
right at $45^\circ$. Differences in the shapes of the curves can be either
due to differences in the properties of the impactor populations or to
differences in target properties. The vertical position of the curve
is a measure of crater density, hence relative age, on the same planet: the
higher the vertical position, the higher the crater density and the
older the surface.
For comparisons amongst different planets, differences in
impact fluxes, impact velocities and target properties must be taken
into account when using crater densities to infer relative ages.

\begin{figure} \centering
\includegraphics[width=\myfigwidthL, angle=0]{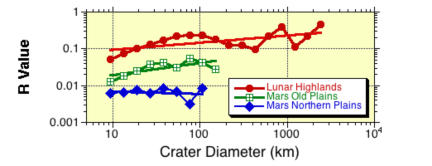}
\caption[]{%
Distorted $R$ plot where the ``Crater Diameter'' axis scale is much larger
than the ``R Value'' axis scale. The undistorted Lunar Highlands and Mars
Northern Plains curves are shown in Figure \protect{\ref{fig:fig01}},
and the undistorted Mars
Old Plains is shown in Figure \protect{\ref{fig:fig09}}.
The solid lines are power law fits to the data. 
}
\label{fig:fig03}
\end{figure}

As with any $\log$-$\log$ plot, 
for the most effective visual communication of data, 
it is good practice to choose
the horizontal (diameter) and vertical
(R value) scales of the plot axes to be the same,
 e.g., $R=0.01$--$0.1$
and $D=10$--$100$ km should have equal lengths, otherwise the curves will
appear distorted. To illustrate, we plot three different crater SFDs
in Figure \ref{fig:fig03} using a longer
$x$ axis than a $y$ axis.  The figure has gross
distortions that make the different populations (Population 1 and 2)
look similar in shape to the fitted straight lines, thereby supressing
statistically significant differences. We note that prior to the
widespread use of computer software graphing programs, $\log$-$\log$ graph
paper was used, which had equal $x$ and $y$ axes scales providing for
undistorted plots. However,  modern computer graphing programs allow
to easily abandon that older paper-based convention, thereby enabling
avoidable degradation of the best possible visual communication of data.
 Log-log plots created with these programs should preferably be
adjusted to display them with equal vertical and horizontal
scales.

\section{Terrestrial Planet Cratering Record}
\label{sec:terrestrial}

\subsection{Population 1 and Population 2}
\label{subsec:pop1pop2}

\subsubsection{Earth}\label{subsubsec:earth}
The Earth is not very useful for reconstructing the impact
cratering record due to its active geological history. The processes
of plate tectonics, deposition and erosion have obliterated most of
its cratering record. About 60{\%} of the Earth's surface (the oceanic
lithosphere) has been destroyed by seafloor spreading during the past
200 million years, or the last 4.5{\%} of Earth history. The average
ocean depth is about 5 km, which would serve to screen out the smaller
impacting objects, somewhat analogous to the screening of impactors by the
Venus atmosphere (c.f.~section~\ref{subsubsec:venus}). 
The continental areas also have been greatly
modified by crustal deformation caused by plate tectonics, and by
erosion and deposition. Only about 180 impact structures have been
confirmed on the Earth. They range in size from $\sim 15$ meters to
$\sim 300$ km
in diameter and have ages ranging from a few years to 2.4 Ga
\citep{French:1998}. Almost all of these craters occur on continental
craton areas. Consequently the crater statistics are not sufficient to
characterize the SFD. There is  no
surface topographic crater record of the very ancient bombardment on Earth because
the solid surface from that period has been almost completely
renewed. However, there are indications of this impact history
preserved in ancient zircon coatings \citep{Trail:2007}, tungsten
isotopes \citep{Willbold:2011}, and layers of impact spherules
caused by large impacts \citep{Johnson:2012}; these are
discussed further in section~\ref{sec:discussion}.

\subsubsection{Moon}\label{subsubsec:moon}
The Moon is the best object in the inner Solar system to
document the ancient cratering record. Unlike Mercury, Mars and Venus,
the heavily cratered lunar highlands have not been greatly modified by
internal or external activity and, therefore, preserve nearly the
entire bombardment record better than any other terrestrial planet.
The Moon's post-mare craters also separately record the geologically recent
impacts.

\begin{figure} \centering
\includegraphics[width=\myfigwidthL, angle=0]{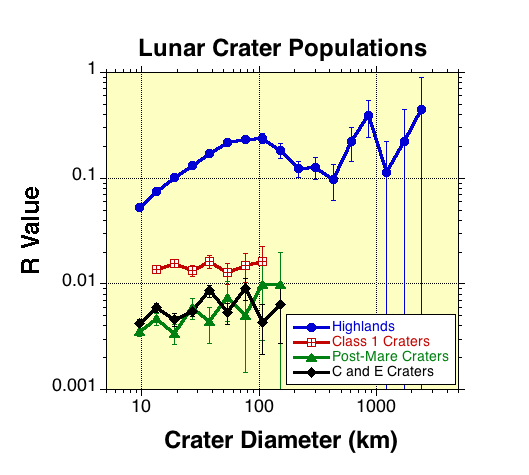}
\caption[]{%
$R$ plot of the two crater populations on the Moon. The top curve (blue)
is for the lunar highlands. 
The middle curve (red) is for all Class 1 lunar craters.
The ``Post-Mare Craters'' (green) are
only those craters that are superposed on the lunar maria, and
the ``C and E Craters'' (black) are the Copernican and Eratosthenian craters
identified stratigraphically as being emplaced during the Moon's youngest
geological period;  all of these are also Class 1 craters.
}
\label{fig:fig04}
\end{figure}

Figure \ref{fig:fig04} shows the SFDs for
four data sets of lunar craters:
lunar highlands craters,
fresh Class 1 craters,
post-mare craters, and
the Copernican and Eratosthenian craters.
The lunar highlands,
Class 1 and the post-mare crater data are from the LPL Crater Catalog
\citep{Arthur:1964};  Copernican
and Eratosthenian crater data are from a catalog by \cite{Wilhelms:1978}.
Class 1 craters are defined as those having a very pristine morphology, a well-defined continuous
ejecta blanket, and fresh secondary craters that post-date the surrounding terrain. They are some of the
youngest craters on a planetary surface \citep{Wood-Anderson:1978}. 
The Copernican and Eratosthenian craters are defined by their stratigraphy;
they are post-mare and they are also all Class 1 in their morphology.
In Figure \ref{fig:fig04}, we have plotted crater data only for $D>8$ km, although the source catalogs do have data at significantly smaller crater diameters.  This lower diameter cut-off is made to avoid confusion with secondary craters; see Section \ref{subsec:popS} for discussion on this point.   We also mention that a modern catalog of lunar craters of diameter $D\ge20$~km, based on data from the Lunar Orbiter Laser Altimeter~\citep{Head:2010}, confirms the SFDs derived from the older imaging data.

As seen in the $R$ plots of Figure \ref{fig:fig04}, the lunar highlands have a high crater density, 
and a complex size frequency distribution; it is the model for the SFD characteristic of Population 1.
The other three curves differ significantly from the highlands' curve:
they have a much lower crater density and they have a 
nearly horizontal straight line shape in the $R$ plot, characteristic of Population 2.  
These represent the cratering that occurred during and/or after a
rapidly declining period of the heavy bombardment.  
The freshest morphological Class 1 craters also have
a size distribution the same as Population 2 craters, 
albeit with poorer statistics, as indicated by their larger 
error bars in Figure \ref{fig:fig04}.

\begin{figure} \centering
\includegraphics[width=\myfigwidthL, angle=0]{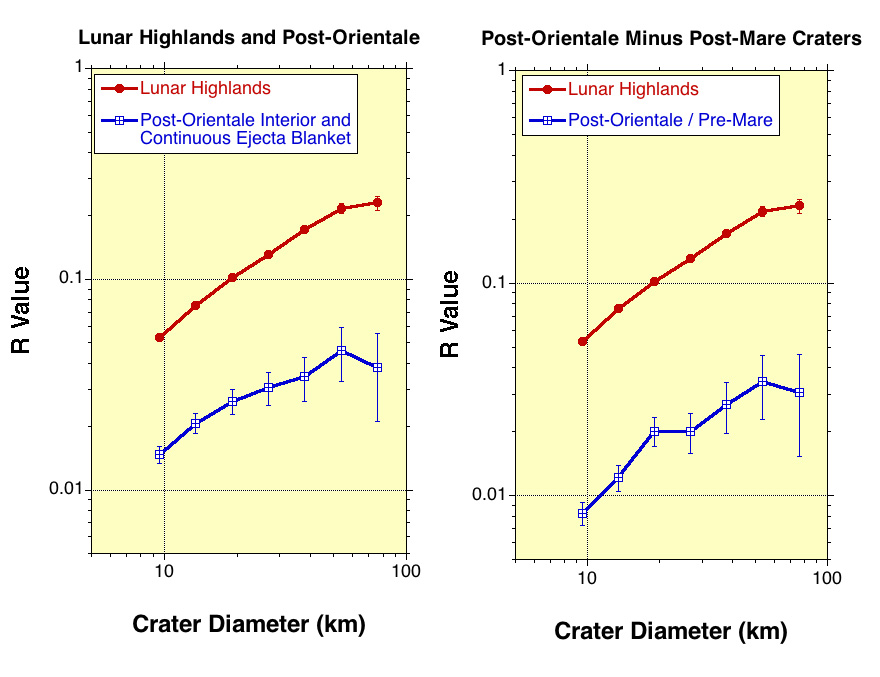}
\caption[]{
$R$ plots of the post-Orientale craters (left) and the post-Orientale minus post-mare craters (right).
The lunar highlands $R$ plot is shown in red for comparison.
}
\label{fig:fig05}
\end{figure}

The freshest large basin on the Moon is the 900 km diameter Orientale basin.
It is thought to be 
the last basin formed by the LHB impactors, with an estimated age of about 3.8--3.7 Ga
\citep[c.f.][]{Lefeuvre:2011}.
Figure \ref{fig:fig05} shows the $R$ plot of newly-determined crater counts on the basin 
interior and the continuous ejecta blanket of Orientale;
these can be called ``post-Orientale'' craters.  (In the Appendix, Figure A-1 shows the imaged area of these counts.) For reference, the $R$ plot of the lunar highlands is also shown in this figure over the same diameter range. 
In the left panel is the post-Orientale curve; this curve slopes down to the
left, but at a gentler slope than the lunar highlands curve. This is
consistent with a mixture of Populations 1 and 2 \citep{Strom:2008}.
In the right panel of Figure \ref{fig:fig05}, the post-mare craters from
a proportional area have been subtracted from the post-Orientale curve to
estimate the Population 1 fraction in the post-Orientale data.
This yields a curve nearly parallel to that of the lunar highlands but
lower by a factor of about 6.5 in the $R$ value.  
Because the Orientale basin may host more Population 2 craters than indicated by just the post-mare crater density, we conclude that at the time of the Orientale basin formation, the Population 1 impact flux had decreased by a factor of at least 6.5 but still dominated over
Population 2.

\citet{Head:2010} have also confirmed the presence of the two crater
populations on the Moon in laser altimetry data obtained by the Lunar
Reconnaissance Orbiter. They have shown that the post-mare crater
population is different from the lunar highlands and that the
Orientale basin has a crater population the same as the highlands,
but at a lower crater density. This was also reported in earlier studies of 
\citet{Strom:2005} and \citet{Marchi:2009}.

\subsubsection{Mercury}\label{subsubsec:mercury}
Mercury has a heavily cratered surface with a widespread
distribution of intercrater plains and a smaller area of relatively
young smooth plains. Previous studies using Mariner 10 data recognized
Population 1 \citep{Strom:1979}, but the statistics of Population 2 were
not good because of the relatively low image quality (i.e., low
resolution and illumination conditions). However, images from the
MESSENGER (MErcury Surface, Space ENviroment, GEochemistry, and Ranging)
data have verified the existence of both Population 1 and
Population 2 craters on Mercury. The largest areas of smooth plains on
Mercury are the Caloris interior and exterior plains and the Northern
Plains \citep{Head:2011}. The MESSENGER mission has now provided the
imaging data necessary to reconstruct more accurately the global
cratering record, especially for the Population 2 craters.
The MESSENGER orbital data is also used to count the freshest craters
[morphological Class 1 from \citet{Arthur:1964,Wood-Anderson:1978}]
on the heavily cratered equatorial areas on Mercury.
These craters are the youngest craters on Mercury and consist of all rayed
craters and those with pristine morphologies and well-developed ejecta
deposits with superposed well-defined secondaries. 
The orbital data is also used for new counts on the Northern plains.

\begin{figure} \centering
\includegraphics[width=\myfigwidthL, angle=0]{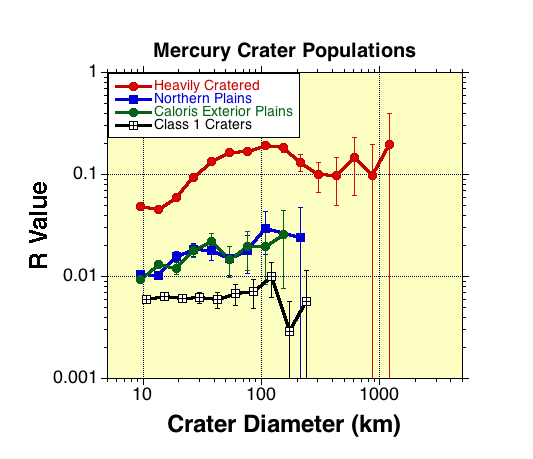}
\caption[]{%
The crater SFDs on several different terrains on Mercury:
the heavily cratered terrains (red), the Northern
Plains (blue), and the Caloris exterior plains (green); the black points
are for all Class 1 craters in the equatorial regions.  
}
\label{fig:fig06}
\end{figure}

Figure \ref{fig:fig06} shows $R$ plots of the crater SFDs on the major geological units of Mercury.
Collectively, the SFD for the heavily cratered terrains with interspersed intercrater plains
(red curve) is similar in shape and magnitude to the lunar highlands shown in
Figure \ref{fig:fig04}.
The upturn at diameters below 10 km is due to secondaries
\citep{Strom:2011}; this is discussed further in section \ref{subsec:popS}.
The green and blue curves are for the Caloris exterior
plains and the Northern Plains. These relatively young plains are the same age
\citep{Head:2011} and have a crater SFD intermediate between Population
1 and Population 2 indicating they are a mixture of the two populations,
but dominated by Population 2 \citep{Strom:2008,Strom:2011}.
Therefore, these plains formed at a time when the impact rate had fallen to a level where Population 2 was beginning to dominate, and well after the Caloris impact.
The black curve is the $R$ plot of
fresh Class 1 craters counted in the equatorial areas.
These fall on an approximately horizontal line characteristic of ``pure'' Population 2.

\begin{figure} \centering
\includegraphics[width=\myfigwidthL, angle=0]{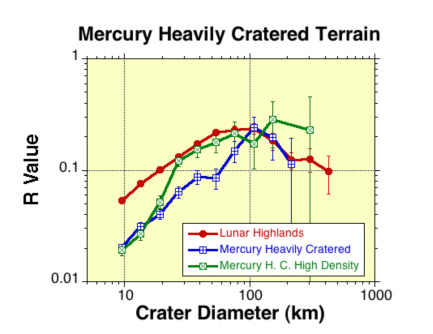}
\caption[]{%
$R$ plots of the crater SFDs on
Mercury's heavily cratered highlands in areas of relatively lower abundance
of inter-crater plains (green points) and in areas of 
abundant inter-crater plains (blue points).  For reference, the lunar highlands
crater SFD is also shown (red points).
}
\label{fig:fig07}
\end{figure}

The widespread intercrater plains of Mercury are older
volcanic plains that are estimated to have been emplaced during the period of the LHB
\citep{Head:2011,Strom:2011}. 
In Figure \ref{fig:fig07} we show the $R$ plots of two high crater density areas of Mercury, one with 
abundant intercrater plains (blue) and one with less abundant intercrater plains (green);  
for comparison, we also plot the lunar highlands (red). We see that the green curve
has a shape and magnitude similar to those of the lunar highlands for crater diameters $D>25$ km,
but at smaller diameters the crater density is lower than that of the lunar highlands; 
this corelates with the existence of intercrater plains near the margin of the high density crater area \citep{Strom:2011}.
We also see that the blue curve (for the area with abundant intercrater plains) has  
lower crater density at diameters all the way up to 100 km. 
Many of the heavily cratered terrains on Mercury show this type of curve \citep{Fassett:2011}.
In a recent analysis, \cite{Marchi:2013} have independently carried out crater counts of Mercury's heavily cratered terrain and reached similar conclusions; additionally they derived an age of the intercrater plains formation beginning at about 4 Ga.

In summary, the global crater counts from MESSENGER images show that Mercury has
been impacted by both Population 1 and 2.  Mercury's Population 1 record has been
affected by the emplacement of intercrater plains while
the largest areas of smooth plains (i.e., Caloris and Northern Plains)
record a mixture of Population 1 and 2.

\subsubsection{Venus}\label{subsubsec:venus}
Venus has undergone multiple global resurfacing events that have
 erased its ancient craters \citep{Strom:1994}.
Its thick atmosphere has progressively screened out smaller objects to
severely modify the crater population below a diameter of about 25 km.
However, the largest craters and multiple craters \citep[c.f.][]{Strom:1994}
provide adequate statistics to reliably give important information on
the geologically recent crater population.

\begin{figure} \centering
\includegraphics[width=\myfigwidthS, angle=0]{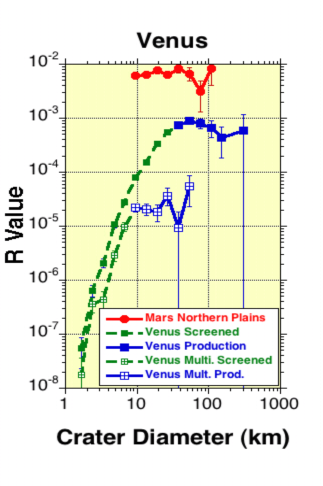}
\caption[]{%
$R$ plots of the crater SFDs of all Venus craters and
Venus' multiple craters (the two sets of blue points). The dashed lines (green) are the portions of the crater SFDs that are estimated to have been
affected by screening in Venus' thick atmosphere; see text for explanation.
For reference, we also show the $R$ plot of the Martian northern plains (red).
}
\label{fig:fig08}
\end{figure}

Figure \ref{fig:fig08} shows the crater SFD on Venus; for comparison, we also show the $R$ plot 
for the Northern Plains of Mars.
For crater diameters above about 25 km, the crater density on Venus is almost an order of magnitude
less than Northern Plains on Mars.
Only very young craters are present on Venus because of
multiple global resurfacing events \citep{Strom:1994}.
At diameters larger than 25 km, the crater SFD has a $-3$ power
law distribution akin to the lunar and martian young Population 2
shown in Figures 4 and 9. 
At crater diameters below 25 km, the curve sharply turns down.
This is because impacting objects of small diameters are severely
affected by atmospheric screening by Venus' thick 90-bar atmosphere
\citep{Zahnle:1992}. 
Part of the Venus crater population consists
of clusters of craters (multiples) that result from fragmentation of
the impacting object when entering the dense atmosphere.
These comprise 16{\%} of all Venus craters. Figure \ref{fig:fig08}
shows the size
distribution of multiples where the diameter is derived from the sum
of the crater areas in the cluster. Multiples are probably formed by
stronger, more consolidated objects that could resist atmospheric
disintegration better than most other impacting objects, but still
weak enough that they broke up in the atmosphere. The turnover of the
curve for multiple craters does not occur until diameters less than
$\sim 9$ km (Figure \ref{fig:fig08}).
At larger diameters the curve is almost flat and
consistent with a Population 2 distribution. This, together with the
much lower crater density, strongly suggests that the impacting
population on Venus was the same as Population 2 on the Moon and
Mars. It is also strong evidence that the turnover of the crater curve
is indeed due to atmospheric screening \citep{Strom:2005}.

\subsubsection{Mars}\label{subsubsec:mars}
\begin{figure} \centering
\includegraphics[width=\myfigwidthL, angle=0]{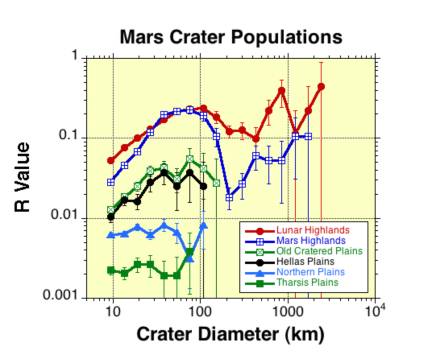}
\caption[]{%
$R$ plots of the crater SFDs on several different terrains on Mars:
 the heavily cratered Mars highlands (dark blue),  two
moderately cratered areas -- Old Cratered plains (green) and Hellas plains (black),
two lightly cratered areas -- the Martian Northern Plains (light
blue) and the Tharsis plains (green).
The Northern Plains are mainly the Vastitas Borealis region including geologic unit
``Mottled Plains Unit'', while the Tharsis plains are the region surrounding the
Tharsis volcanic constructs including the geologic unit ``Volcanic
Plains on Tharsis Montes Region''. The Hellas Plains are the plains
within the Hellas basin. The Old Cratered Plains are the plains east of Tharsis that partly includes the geologic unit
``Old Volcanic Material''. %
}
\label{fig:fig09}
\end{figure}

Mars provides an excellent record of the inner Solar system
cratering history. It has numerous large areas with a wide variety of
ages that provide good crater statistics in several geological periods
spanning almost the entire
history of the Solar system. Although some of the heavily cratered highlands
have been greatly modified by internal processes and erosion and
deposition, some portions of the relatively old areas show Population 1
very well, though at a lower crater density than the most heavily
cratered areas. Furthermore, the large areas of younger plains show a
considerable variety of ages that provide good counting statistics for
determining the geologically recent cratering record. Figure \ref{fig:fig09} is a
crater SFD of the different crater populations on
Mars compared with the lunar highlands. (The various geologic units of Mars
are as defined in the Geologic Map of Mars by \citet{Scott:1978}.)
In contrast with the lunar
highlands, the Mars highlands curve shows a noticeable depletion of craters at
diameters below about 30 km.  
From the geological context, this is due to erosion and deposition.
The Old Cratered Plains $R$ plot is for the surface unit east of Tharsis while the Hellas
plains $R$ plot is for the surface unit within and surrounding the large Hellas impact basin.
Both of these show the characteristic Population 1 shape 
albeit at lower density than the highlands.
There are a number of young lightly cratered plains on Mars.
Two representative examples, the northern plains and
the Tharsis plains, are shown in Figure \ref{fig:fig09}.
Both have a horizontal straight line $R$ plot, characteristic of the Population 2
craters. In general, Mars displays the two crater
populations and their transition better than other terrestrial planets
because it has large surface areas with a variety of crater densities hence a variety of relative ages.

\subsection{Population S--the Secondary Crater Problem}
\label{subsec:popS}
A distinct population of impact craters on planetary surfaces is caused
by ejecta from primary impacts \citep[e.g.][]{Shoemaker:1965,McEwen:2005}.
The size and spatial distribution of secondary craters (Population S)
generally depend on the size of the primary impact crater, the impact
velocity, and the planet or satellite gravity field \citep[e.g.][]{Xiao:2014}.
At small diameters, secondary craters outnumber primaries
on terrestrial planets and the Moon by orders of magnitude because a
single primary impact can produce thousands of secondary craters
\citep[e.g.][]{Dundas-McEwen:2007}. Large basins such as Orientale and
Imbrium on the Moon have produced some secondaries up to 20 km in
diameter. However, at such large diameters,
 their number is relatively small compared to
primaries in heavily cratered terrain. Our $R$ plots for heavily
cratered terrain have a lower diameter cut-off of 8 km to 
avoid confusion with the vast majority of basin secondaries.

\begin{figure} \centering
\includegraphics[width=\myfigwidthL, angle=0]{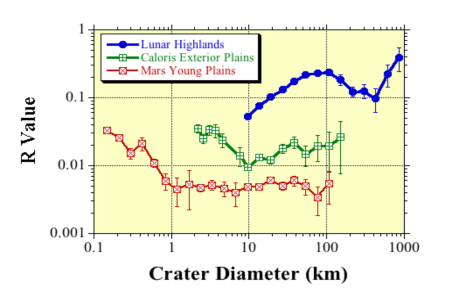}
\caption[]{%
$R$ plots for the SFD on Mercury's Caloris Exterior Plains and
Mars Young Plains, down to small crater sizes where
 secondary craters are abundant. 
On Mercury the upturn in the curve (green) at diameters below 10 km is dominantly secondary craters,
while on Mars (red) the secondary upturn occurs below 1 km.}
\label{fig:fig10}
\end{figure}

Secondary impact craters are very widely distributed and dominate the
small crater population on planetary surfaces.
\citet{Robbins:2011} have shown that secondaries on Mars are
very widespread
and can affect crater age dating unless they can be unambiguously
distinguished from primaries. \citet{Xiao-Strom:2012} showed that
secondaries dominate the small crater ($D<1$ km) population on 
both young and old lunar surfaces. 
The secondary craters' SFD is characterized by a $-4$ 
power law.
Figure \ref{fig:fig10} shows three $R$ plots in which the 
dominant presence of secondary craters is evident in the sharp change
of slope at small diameters.  This upturn in the $R$ plots occurs at crater
diameters below about 1 km on Mars. 
On the Moon secondaries can begin to affect the crater SFD at diameters less than about 1 km, but on Mercury
 this occurs at a diameter of about 10 km.  This is the main reason that
 most of the $R$ plots in this paper have a lower cut off diameter of 8 km.

On the Moon, the basin secondaries mapped by \citet{Wilhelms:1978} 
account for only 15\% of the craters in the 8--11.3 km diameter bin, and even
less at larger diameter bins.
However, on Mercury the secondaries are larger than on any other terrestrial planet. 
The Mercury heavily cratered terrain as well as the Caloris exteroir plains $R$ plots have an
upturn in the curve in the 8--11.3 km size bin that is due to secondaries
as shown in Figures 10 and 6.  The widespread distribution of larger
secondaries on Mercury may be due, at least in part, to a combination
of higher secondary impact velocities, higher ejection angles and the
larger surface gravity on Mercury \citep{Strom:2008,Strom:2011,Xiao:2014}.

\begin{figure} \centering
\includegraphics[width=\myfigwidthL, angle=0]{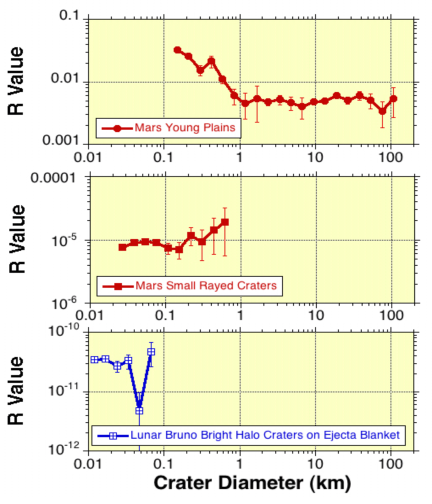}
\caption[]{%
$R$ plots for the crater SFDs of 
Mars Young Plains down to sub-km diameter sizes (top panel),
 Martian small rayed craters (middle panel), 
and bright haloed craters on the continuous ejecta blanket of
the young and very fresh Giordano Bruno crater on the Moon (bottom panel).}
\label{fig:fig11}
\end{figure}

The upturn in the crater SFDs at small
diameters has been interpreted by some authors to be the primary
crater production function
\citep[e.g.][]{Hartmann:2007,Ivanov:2002,Michael-Neukum:2010}.
This is very unlikely as demonstrated by
\cite{McEwen:2006,Xiao-Strom:2012}.
Direct evidence supporting this statement is found in the SFDs of craters
with rays and bright halos on inner Solar system bodies which are most
likely primaries.  
In Figure \ref{fig:fig11}, we show the $R$ plots of Mars Young Plains craters
down to small crater sizes, compared with two populations
of rayed craters of small diameters, from about 1 km down to about 10 meters. 
The middle panel in the figure shows the $R$ plot of the small rayed crater
population on Mars, and the ``Lunar Bruno'' curve in the bottom panel
is for the bright haloed craters on the continuous ejecta blanket of the 
very young Giordano Bruno crater on the Moon. 
The Mars rayed craters show a gentle slope upward to the right in the two
largest diameters but the statistics are poor and a $-3$ differential SFD is well
within the error bars.  
The gentle decrease in the number of rayed craters at diameters less than 0.02 km 
is likely due to the loss of rays at decreasing crater size, analogous to the loss
 of multiple craters below $D\approx 8$ km on Venus due atmospheric 
 screening (cf.~Figure \ref{fig:fig08}).
If there was a loss of rays throughout the diameter range counted then 
a systematic downward trend over the entire diameter range would occur,
which is not present. 
This supports the interpretation that the upturn seen at small
diameters, $D\lesssim1$ km, in the $R$ plot of the Mars Young Plains
craters (see Figure \ref{fig:fig10}) as well as in other 
cratered terrains on Mars, is not reflecting the primary crater population, 
but is owed to large numbers of secondary craters in that size range.
Likewise, the lunar Bruno crater curve (bottom panel of Figure \ref{fig:fig11}) 
is flat in the $R$ plot down to diameters of $\sim 10$ meters, indicating that
this small diameter primary crater population at geologically recent times
on the Moon also shares the differential $-3$ slope characteristic of Population 2
craters at larger sizes.

Some crater counters believe they can distinguish between primaries
and secondaries, but this is doubtful unless they have and use
topographic data to determine their depth-to-diameter ratio.  This
technique is unlikely to be applied to every small crater due to their
enormous numbers. Eliminating crater clusters or strings does not
eliminate all secondaries because much ejecta are on high trajectories
that produce randomly distributed craters, i.e., distant secondaries
\citep{Xiao-Strom:2012}. These secondaries usually do not occur in
clusters or chains, and may be highly circular in shape similar to
same-sized primaries, making them difficult to distinguish from
primaries. On Mercury, the contamination of secondaries might be
extremely severe because some craters form very circular and isolated
secondaries, even on continuous secondaries facies, probably due to
the special target properties \citep{Xiao:2014}. 
An empirical way to
make the distinction between primaries and secondaries is to plot the
crater counts on an $R$ plot. If the curve trends upward to the left
at small diameters ($D\lesssim 1$ km for Mars and lunar craters, $D\lesssim 10$ km
for Mercury craters),  then the count is likely contaminated with
secondaries.

\subsection{Summary of the Inner Solar System Cratering Record}
\label{subsec:innersummary}

\begin{figure} \centering
\includegraphics[width=\myfigwidthL, angle=0]{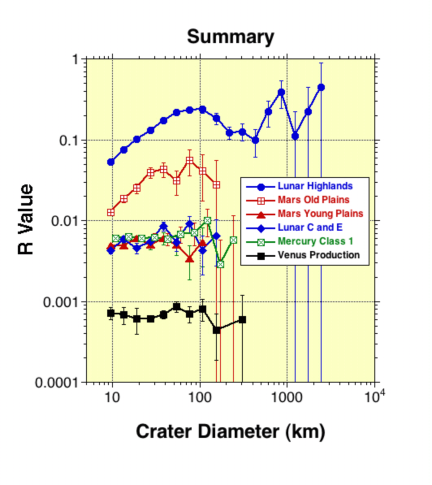}
\caption[]{%
These $R$ plots summarize the inner solar system cratering record
for crater sizes in the range of about 10 km to about 1000 km.
They show two distinctly different crater populations.
The curves above an $R$ value of about 0.01 have a complex shape
characteristic of Population 1, and the lower curves have a nearly
horizontal straight line shape chacteristic of Population 2. 
 The ``Lunar C and E'' craters are post-mare Copernican and
Eratosthenian in age, and the ``Venus Production'' is a composite of
the productions of all craters and multiple craters
(see Figure \protect{\ref{fig:fig08}}).
}
\label{fig:fig12}
\end{figure}

Based on their different crater SFDs,
the terrestrial planets and the Moon have been impacted by two populations
of objects: Population 1 dominated at early times and was associated with a much higher impactor flux than Population 2 which dominated at later times but beginning at least as early as the formation of lunar maria and up to the present time.
Figure \ref{fig:fig12} summarizes the two crater populations in the inner
Solar system from the heavily cratered lunar highlands, the martian
old cratered plains, and the younger more lightly cratered plains on
the Moon, Mars, and Venus. Population 1 is responsible for the period
of Late Heavy Bombardment, and Population 2 is responsible for the
period mostly after heavy bombardment up to the present time.
The Venus curve is a composite of the production population for all
craters and for multiple craters only. The Mercury curve is for Class 1
craters, but Mercury's Northern Plains and the Caloris interior and
exterior plains are mixtures of Populations 1 and 2
(see Figure \ref{fig:fig06}).
Also, large numbers of secondary craters dominate the small crater
populations on these bodies; in the $R$ plots, the secondaries contamination is signalled by a distinct upturn for diameters $D\lesssim1$ km on the Moon and Mars, $D\lesssim10$ km on Mercury.

\subsection{The Outer Solar System Cratering Record}
\label{subsec:outercrater}

\begin{figure} \centering
\includegraphics[width=\myfigwidthL, angle=0]{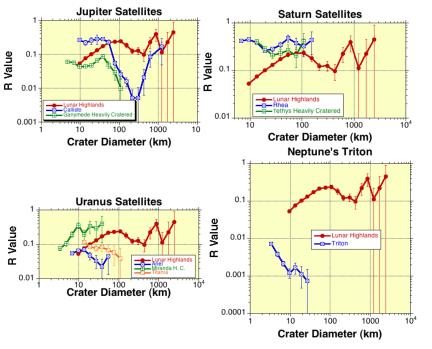}
\caption[]{%
The impact crater SFDs on the heavily cratered surface
units of the outer planet satellites.  In each panel, for reference we also 
show the $R$ plot for the lunar highlands craters. 
}
\label{fig:fig13}
\end{figure}

The cratering record on outer solar system satellites appears to be
very different from that in the inner solar system
\citep{Chapman:1986,Dones:2009,Strom:1981,Strom:1990,McKinnon:1991}.
Figure \ref{fig:fig13} is a group of $R$ plots
of the SFD of impact craters on the satellites
of Jupiter, Saturn, Uranus and Neptune compared to the lunar
highlands. Only satellites with heavily cratered surfaces are shown in
the plots. These data show that, with the possible exception of the heavily
cratered surface of Miranda, the crater SFDs of the
satellites are different from those of the lunar highlands Population 1
craters. On the Uranus satellites Ariel and Titania, at smaller
crater sizes the curves slope upward compared to the downward slope on
the lunar highlands, but on the heavily cratered terrain on Miranda
the curve slopes downward similar to the lunar highlands.
On Triton the curve slopes upward at a steep angle as we go to smaller
crater sizes. However, this satellite has been greatly resurfaced and
these craters represent late impacts probably well after the period of
late heavy bombardment. None of the outer planet satellites' crater
populations resemble the cratering record on the heavily cratered
terrain of the Moon and the inner planets. Therefore, Population 1
craters appear to be confined to the inner solar system.
It is possible that the outer planet satellites may have been impacted
by a mixture of projectile populations comprised of both comets and
planetocentric objects.  However, this topic requires further study
and is beyond the scope of the present paper.

\section{Discussion}
\label{sec:discussion}

\subsection{Sources of Population 1 and Population 2 Impactors}
\label{sec:sourcepop12}

\begin{figure} \centering
\includegraphics[width=\myfigwidthL, angle=0]{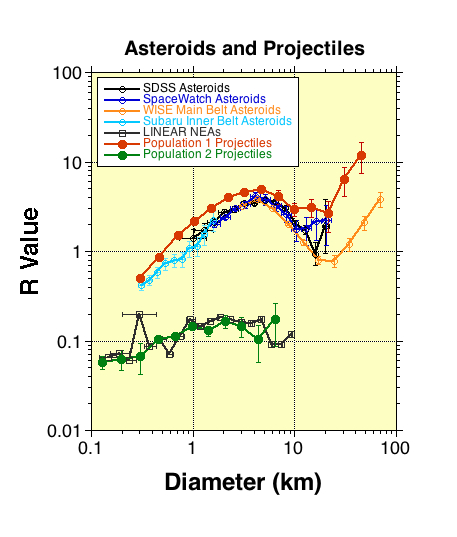}
\caption[]{%
The SFDs of the impactors derived from the
crater SFDs compared with those of Main Belt
Asteroids and Near Earth Objects. The red 
points are impactors derived from the lunar highland crater
distribution (Population 1; Figure \protect{\ref{fig:fig01}}),
and the green  points are
derived from the young Mars plains crater population (Population 2;
Figure \protect{\ref{fig:fig01}}).
The vertical positions of the asteroid R-plots are arbitrary;
the scale factor is chosen for clarity of comparison. See text for
detailed explanation.
}
\label{fig:fig14}
\end{figure}

The size of an impact crater is related to the size of the impactor by
the Pi-group crater scaling law
\citep{Croft:1985,Schmidt:1987,Melosh-book:1989,Collins:2005}.
\citet{Strom:2005} adopted this
procedure to obtain the SFDs of the impactors
responsible for the Population 1 and Population 2 craters.
We follow the same procedure here; specifically, we use the web-based
calculator of \citet{Melosh-Beyer:1999} to compute the impactor sizes
for each of the crater size bins.
We assume a target type of ``competent rock'', adopt a common target
and projectile density of 3000 kg ${\rm m}^{-3}$, a single
impact angle of 45 degrees; we adopt surface gravity of 1.6 m ${\rm s}^{-2}$
and 3.7 m ${\rm s}^{-2}$ for the Moon and Mars, respectively.
The lunar highlands
craters were adopted as best representing Population 1 because the
lunar highlands are the least affected by geological activity.
The Martian young plains were adopted as best representing Population 2
for their better statistics compared with young crater populations on
the other terrestrial bodies.  For simplicity, a single value of the
characteristic impact velocity was adopted for each of these (18.9
km/s for the Moon, and 12.4 km/s for Mars). These values are the
median impact velocities obtained in a recent self-consistent
dynamical model of asteroid impacts on the terrestrial planets
\citep{Minton:2010}.
The resulting impactor SFDs are shown
in Figure \ref{fig:fig14}.
Slightly different median impact velocities than adopted
here can be found in the literature (e.g., \citet{Ivanov:2002} obtain
17 km/s for the median velocity of asteroid impacts on the Moon,
\citet{Bottke:2012} quote values as high as 21 km/s for dynamical
models of ancient asteroid bombardment on the Moon).
If we adopt impact velocities that are smaller or larger by 20\% than
the nominal value adopted here (i.e., 15.1 km/s or 22.7 km/s),
the peak of the lunar highlands impactors' curve shifts to $\sim 0.5$ km
larger or smaller impactor diameter, respectively,
but the shape of the distribution does not change.

Also plotted in Figure \ref{fig:fig14} is the available data on the size
distributions of the near Earth objects (NEOs) and the main belt
asteroids (MBAs).  The NEO size distribution is based on the
bias-corrected LINEAR diameters for NEOs \citep{Stuart:2004}.
The MBAs size distribution is based on four published data sets:
1) Spacewatch \citep{Jedicke:1998};
2) Sloan Digital Sky Survey (SDSS; \citet{Ivezic:2001});
3) Wide-field Infrared Survey Explorer (WISE) first release data~\citep{Masiero:2011}
and
4) Subaru Main Belt Asteroid Survey \citep{Yoshida:2003}.
From the SDSS data set, we used the ``red'' asteroids data, and from the other data sets 
 we used only the data on the inner part of the main asteroid belt.

The Spacewatch survey did not measure albedos nor determine taxonomic types.
In order to convert their asteroid absolute magnitude data to
asteroid diameters, we used weighted average albedo values that were
calculated as follows.
(a) We used the Spacewatch dataset of absolute
magnitude $(H)$ from Table IV of \citet[][p.~256]{Jedicke:1998}.
This table partitions the main belt into three parts, inner, middle,
and outer, and gives binned statistics for the numbers of asteroids in
0.5 magnitude-wide bins.
(b) We used the results of \citet{Yoshida:2003}
for the statistical ratio of the relative abundance of
different taxonomic types (S- or C-type asteroids) in each part of the
main belt. \citet{Yoshida:2003} carried out a deep color survey of
the main asteroid belt using the Subaru telescope; in this work,
they obtained the colors and approximate orbital locations
(inner/middle/outer belt) of 861 MBAs, and estimated the relative
abundance ratios of S- and C-types in the inner/middle/outer main belt.
(c) We calculated the average albedo values, $\left< A \right>$,
for each taxonomic type from the catalog of asteroid albedo and taxonomic
types in the Planetary Data System catalog%
\footnote{%
http://sbn.psi.edu/pds/archive/physical.html
}.
We found
  $\left< A \right> = 0.21$ for S-type, and
  $\left< A \right> = 0.05$ for C-type.
These values are consistent
with recent estimates from infrared space surveys such as WISE and AKARI
\citep{Masiero:2011,Usui:2013}. (d) Finally, we calculated a weighted average albedo
for each partition (inner/middle/outer) of the asteroid belt.
For example, for the inner MBAs, Yoshida found 233 S-type and 112 C-types,
so the weighted average albedo for inner belt MBAs is
$(0.21 \times 233 + 0.05 \times 112) / (233+112) = 0.158$.

In Figure \ref{fig:fig14}
the derived impactor SFDs of Population 1
and Population 2, as well as the data for the MBAs and NEOs are shown
in an $R$ plot. The vertical position of the asteroid curves is arbitrary; these
positions were chosen to clearly compare the shapes of the various
curves. This figure differs slightly from a similar one published in
\citet{Strom:2005} in that the larger lunar basins have been included in the
Population 1 curve and we have also included the WISE data for MBAs. 
Although the impact mechanics for large basins are
not as well understood as for smaller craters 
 (and old lunar basins may have 
formed in a hot crust, which may also affect the final crater size),
the size bins at these diameters are so large (several hundred km) that the uncertainty in
the derived projectile mean diameter is probably within the size bin.

We see that the Population 1 impactor curve is very well matched with
the curve for the MBAs.  Only the largest size bin representing the three largest lunar basins (Imbrium, South Pole-Aitken, and Procellarum) is significantly below the MBAs data point.
This may be due to the unrecognized structure of other large basins
that formed early in the period of LHB and were obliterated by the heavy
bombardment. \citet{Feldman:2002} have found geochemical signatures
of large basins on the Moon's far side that are similar to that of
the Procellarum basin. These are at least part of the missing basins.

There are two conclusions that can be drawn from the
impactors/asteroids comparison in Figure \ref{fig:fig14}.
First, the size distribution of MBAs is virtually identical to the size distribution
of Population 1 impactors; this was also observed by
\citet{Neukum:2001}. This result indicates that the Population 1 impactors
originated from Main Belt Asteroids or possibly a population that had
the same size-distribution as the contemporary inner main asteroid
belt. Second, the comparison of LINEAR data and the young crater
Population 2 strongly indicates that Population 2 craters were made by
impactors derived primarily from Near Earth Objects. Supporting
evidence that inner Solar system impactors were asteroids rather than
comets is found in trace element analyses of lunar samples returned
during the Apollo program \citep{Kring:2002}. Furthermore, direct
fragments of impactors have been identified in a recent study of
ancient ($>3.4$ Ga) and younger ($<3.4$ Ga) lunar regolith samples;
these show that lunar impactors were primitive chondritic asteroids prior
to $\sim 3.4$ Ga, but the younger impactors have more diverse chemical
compositions \citep{Joy:2012}.

Let us consider in some detail the case of Population 1 impactors.
Many previous studies have held that the ancient craters were made by
a declining population of planetesimals in the inner Solar system that
were left-over from planet formation.  However, such a source is
untenable because the typical dynamical lifetimes of planetesimals in
planet-crossing orbits in the inner Solar system are $<10^7$ years
\citep{Gladman:1997,Ito-Malhotra:2006}. When collisions are
taken into account, the lifetime of an inner Solar system left-over
planetesimal population is reduced even further \citep{Bottke:2007}.
The only known long-lived population that is a viable source of the
Population 1 impactors is the main asteroid belt (but see discussion
below about the Hungaria asteroids).  A main asteroid belt source is
consistent with the close match between the old Population 1 impactors
and the contemporary MBAs, provided that (a) the shape of the MBAs'
size-frequency distribution achieved a steady-state at least as early
as $\sim 4$ Ga, and has remained nearly unchanged since then, and (b) a
dynamical mechanism existed at ancient times for transporting main
belt asteroids into planet crossing orbits in a size-independent
way. We discuss the latter condition in detail in Section~\ref{subsec:LHBdynamics}.
For the former condition, we note that numerical modeling studies by
\citet{Cheng:2004} and \citet{Bottke:2005} of the collisional evolution of
the asteroid belt find that its size distribution changes little after the
first $\sim 100$ Myr.

\begin{figure} \centering
\includegraphics[width=\myfigwidthL, angle=0]{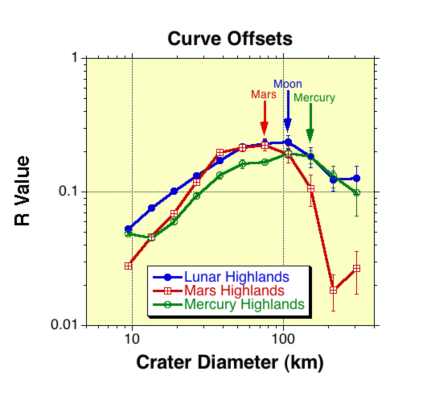}
\caption[]{%
Displacement of the crater SFD curves 
attributed to differences in the median asteroid impact velocities at
Mercury, the Moon and Mars. The arrows indicate the
``downturn'' location of each curve.  The locations of the
peaks are also similarly displaced.  See text for detailed explanation.}
\label{fig:fig15}
\end{figure}

There is additional evidence in the crater record
supporting the hypothesis that the main asteroid belt was the primary
source of the Population 1 impactors. This evidence lies in a
comparison of the crater size distributions of the heavily cratered
terrain of Mars, the Moon, and Mercury at the larger diameters where
the curves have significant downturns to steeper slopes (i.e.,
more negative power law index; Figure \ref{fig:fig15}).
The downturn in the Mercury crater curve 
occurs at a larger diameter size-bin than on the Moon,  
whereas on Mars, the downturn occurs at a smaller diameter size-bin, 
as indicated by the arrows in Figure 15. 
We estimated the location of the peak of each of the $R$ plots 
in Figure \ref{fig:fig15} by fitting a 5th order polynomial to each curve;
the peaks in the best-fit polynomials occur at crater diameters of 
 $99^{+7}_{-5}$~km, $81^{+5}_{-3}$~km and $69^{+1}_{-5}$ km
for Mercury, Moon and Mars highlands, respectively.  
This systematic shift from Mars to the Moon to Mercury of
the ``downturn diameter''  of large craters is consistent with an
origin of impactors from the Main Asteroid Belt, because the median
impact velocities of these asteroids are higher on Mercury and lower
on Mars, compared to the Moon. 
Use of the Pi-group scaling law and adopting the median impact velocity of MBAs 
for each planet (38.1 km/s, 18.9 km/s and 12.4 km/s for Mercury, Earth-Moon, and Mars respectively \citep{Minton:2010}), yields corresponding peak impactor sizes of 
 $3.1^{+0.2}_{-0.2}$, $3.4^{+0.3}_{-0.1}$ and $3.9^{+0.05}_{-0.3}$ km, 
respectively.  Within the $\pm0.5$~km uncertainty owed to the uncertainty in the mean impact velocities,  these peak diameters of the Population 1 impactors on Mercury, Moon and Mars are the same. They are also the same as the local peak (near $D\sim$~3--4~km) of the $R$ plot of the MBAs size distribution (cf.~Figure \ref{fig:fig14}).
In other words, for the same impactor size distribution,
the systematic differences in the mean impact velocity at Mercury, Earth-Moon and Mars, 
produce a shift in the crater sizes that
are consistent with the observed shifts in the crater size
distributions of Population 1 on Mercury, Moon and Mars. 
These shifts are therefore consistent with the hypothesis
that the objects responsible for the Population 1 craters originated
directly in the Main Asteroid Belt. This shift and its implication for
the orbits of the impacting objects were first noticed by
\citet{Strom:1988}; its significance
and connection with the Late Heavy Bombardment 
was explained by \cite{Malhotra:2011}.

This shift also indicates that Mercury Population 1 was unlikely to
have been due to Vulcanoids interior to Mercury's orbit
\citep{Stern:2000} because Vulcanoids would have impacted Mercury at about
13--14 km/s, similar to the impact velocity of asteroids at Mars.
In this case the curve would show an offset similar to that of Mars;
this is not observed. Also, if Vulcanoids existed, they may not be the main
impactor-source for the Population 2 craters on Mercury, unless they
had the same SFD as the NEOs. The MESSENGER
spacecraft has not yet discovered any candidate Vulcanoids,
indicating this hypothesized asteroid belt may have been depleted
if it once existed.

On the other hand, with regard to the Population 2 craters,
Figure \ref{fig:fig14}
shows that the SFD of projectiles responsible for these
is quite different from that of the Population 1 impactors.
The differences are illustrated by the value of the asymptotic slope of
the power law SFD at small diameters, $D<2$ km:
the Population 2 impactors have a $-2.8$ asymptotic slope that is
significantly steeper than the Population 1 impactors' $-2.2$ slope.
Moreover, Figure \ref{fig:fig14} also shows that the SFD of
the Population 2 impactors is very similar to that of the NEOs.  This is perhaps the most direct evidence that the source of
the Population 2 impactors is the NEOs.
However, this conclusion raises a number of issues that we discuss below.

The NEOs are a transient population, with typical
dynamical lifetimes $\sim 10^7$ years whereas Population 2 craters have
accumulated over more than $\sim 3$ gigayears.
Indeed \citet{Lefeuvre:2011} show that the density of Population 2 craters
on the Moon is consistent with a nearly constant impact flux similar to that
of the contemporary NEO impact flux over the past $\sim 3.5$ gigayears;
\citet{Grieve:1994} and \citet{Neukum:1994HDCA} had
previously reached a similar conclusion\footnote{
It is possible that the flux has varied by a factor of two or three over the past $\sim3$ Gyr \citep{Hartmann:2007, Marchi:2009, Kirchoff:2013}.
The crater record that we are examining is integrated over any variations in the flux that may have occurred and does not affect the conclusion that Population 2 is from NEOs.}.
How can this be reconciled
with the short dynamical lifetimes of NEOs? Quite independent of the
recently discovered similarity of the Population 2 craters and the
NEOs' size distributions, previous dynamical studies of asteroids
have indicated that the transient population of NEOs can be maintained
in nearly steady state over gigayear timescales by being resupplied
primarily from the main asteroid belt [see review by \citet{Morbidelli:2002}].
Importantly, the steeper size distribution of NEOs
compared to that of the main asteroid belt indicates that the
dynamical transport process must be size-dependent,
favoring the injection of smaller asteroids into the inner Solar system.
A number of studies have argued that the key ingredient is the effect of
non-gravitational forces owed to thermal radiation, specifically the
Yarkovsky and YORP effects; \citet{Bottke:2006} provide a recent review.

The Yarkovsky effect, named for the Polish engineer who
discovered it more than a century ago, is a small thermal thrust that is
produced when small airless spinning bodies orbiting the Sun emit
thermal radiation in equilibrium with absorbed sunlight but with a
small delay owed to thermal inertia; this small thrust causes a net
secular orbital drift that depends on the size and spin and material
properties of the body. The same physical process also produces a
torque that modifies the small body's spin rate and spin axis
orientation, and is referred to as the YORP
(Yarkovsky-O'Keefe-Radzievskii-Paddack) effect \citep{Rubincam:1988}.
\citet{Farinella:1999} showed that over a few tens of millions of
years these effects are large enough to push a significant number of
sub-20-km size asteroids into strong Jovian resonances;
the latter then deliver them into terrestrial planet-crossing orbits
and thereby into the NEO population. The Yarkovsky effect and the YORP
effect are most significant for objects between 10 cm and 10 km diameter;
both effects diminish significantly beyond this size range.
\citet{Morbidelli:2003a} numerically modeled the dynamical origin of NEOs
from MBAs, finding that, under a plausible range of adopted model
parameters, the combination of collisions and the Yarkovsky and YORP
effects roughly explains the steeper size distribution of the NEOs
compared with the SFD of their source, the main asteroid
belt. Further detailed studies are needed to determine whether the
difference between the size distribution of the NEOs and MBAs is
quantitatively fully accounted for by these non-gravitational effects,
or whether this difference hides additional surprises.

Regardless of the reasons for the difference between the size
distributions of the NEOs and of the main asteroid belt,
the conclusion that Population 2 impactors' size distribution is similar
to that of the NEOs holds.  

It is of some interest to note that recent studies of the spatial
distribution of young craters on the lunar surface find a significant
longitudinal asymmetry due to the Moon's synchronous rotation
\citep{Morota:2003}.  The magnitude of this asymmetry is roughly
consistent with the NEOs being the impactors
\citep{Gallant:2009,Ito-Malhotra:2010,Lefeuvre:2011}.
However, \citet{Ito-Malhotra:2010}
note a small discrepancy between the observations
and the theoretical model and suggest that it may indicate a missing
tail of low velocity Earth-Moon impactors~\cite{JeongAhn:2010}.

\subsection{Age and Duration of the Late Heavy Bombardment}
\label{subsec:LHBage}

The existence and properties of the two crater populations 
support the hypothesis of a ``terminal lunar cataclysm'', and, more widely, that the Late
Heavy Bombardment (LHB) was a spike in the impact flux common to all
the terrestrial planets, and that the spike consisted of bombardment by
Population 1 projectiles whereas the post-spike projectiles have been
Population 2.  However, these impact crater data on their
own do not constrain the timing and duration of the LHB.
The latter are obtained from laboratory analysis of lunar samples
and meteorites.

The analyses of samples from the Apollo lunar program showed that the lunar crust is $\sim 4.5$ Gyr old \citep{Tera:1973,Norman:2003}, but that several hundred million years subsequent to differentiation and crust formation, the lunar highlands suffered extensive mobilization of Pb isotopes and widespread impact metamorphism over a relatively short time interval ($\sim200$ myr) that ended $\sim 3.8$ Ga \citep{Turner:1973,Tera:1974}.  This first led to the ``lunar cataclysm'' hypothesis of a spike in the bombardment at $\sim 3.9$ Ga.  

The ages of the large lunar basins also apparently cluster near 3.9 Ga \citep{Turner:1973,Ryder:2002}.   
Also, the Martian meteorite Allan Hills 84001 records a shock event at 3.92 Ga \citep{Turner:1997}.  These disparate pieces of evidence suggest a spike in the impact flux in the inner Solar system several hundred million years after the formation of the planets.

On Earth, there is possible evidence of the LHB recorded by impact
generated metamorphic over-growths on zircons older than $\sim 3.5$ Ga.
The ages of the over-growths cluster at $\sim 3.9$ Ga and may be due to
multiple impact events associated with the LHB \citep{Trail:2007}.
\cite{Willbold:2011} report that analysis of $\sim3.8$ billion-year-old rocks from Isua,
Greenland revealed a significantly higher isotopic tungsten ratio
${}^{182}W/{}^{184}W$ than modern terrestrial samples; they suggest that the Late Heavy
Bombardment may have triggered the onset of the current style of mantle convection on the Earth.  
However, interpretation of the terrestrial Hadean eon record remains highly uncertain due to the complex
geological history of our planet.

On Mars, \citet{Frey:2008} has identified old impact basins that cluster around crater-density-based ages of about 4.2--4.1 Ga.  However, these crater ages may be an overestimate as they are based on the assumption that there was a smooth decline in the impact rate since the origin of the solar system, $\sim4.5$ Ga. 
 
The exact onset age and the duration of the LHB are a subject of current debate~(e.g., \cite{Chapman:2007}).  
In a recent paper, \cite{Norman:2014} report a large basin-scale melting event on the Moon at $4.22\pm0.01$ Ga, based on new measurements of U-Pb isotopic compositions in a lunar melt rock sample; they suggest an earlier onset of the basin-forming epoch that was more prolonged and less intense than inferred from previous lunar sample studies.   
This interpretation of the impact chronology of the inner solar system attributes the concentration of lunar highland impact melt and breccia ages at about 3.9--3.7 Ga to a sampling bias. 

The possibility that the Apollo lunar samples suffer from a sampling bias and reflect the age of a single large basin-forming impact, the Imbrium basin, has been discussed in the lunar literature~(e.g., \cite{Haskin:1998}).  
A recent study of zircons from the Apollo 12 landing site finds that the Imbrium impact occurred $3.92 \pm0.013$ Ga \citep{Liu:2012}. 
\cite{Stoffler:2006} reviewed the radiometric ages of Apollo samples and compiled a list of the ages of lunar highlands impact breccias and melts, including clast-poor impact melts (10 samples), crystalline melt breccias (21 samples), fragmental breccias (3 samples) and granulitic breccias and granulites (10 samples).  All of these ages lie between 3.7 and 4.2 Gy.  We note that, of the 45 samples' ages listed by \cite{Stoffler:2006}, 29 (64\%) have error bars outside of the Imbrium impact age of 3.92 Gyr. Most of these (90\%) are less than the age of the Imbrium impact (see Figure A-2 in the Appendix). 
These data indicate that most of the lunar impact breccia and melts are not related to the Imbrium impact.

\begin{figure}[!ht] \centering
\includegraphics[width=\myfigwidthS, angle=0]{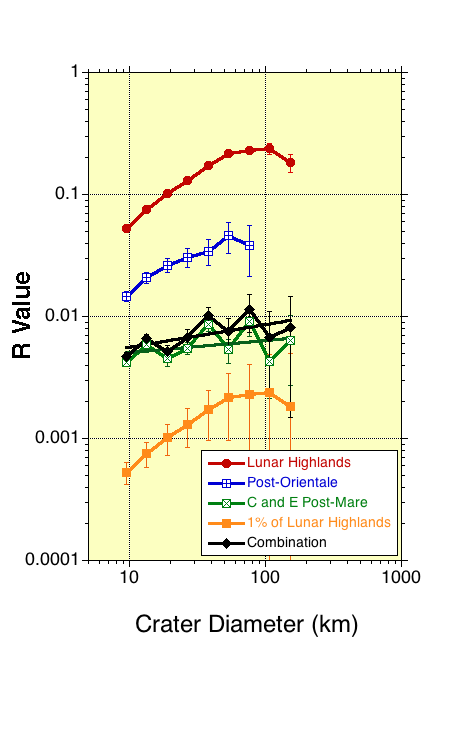}
\caption[]{%
The top curve (red) shows the lunar highlands $R$ plot and the blue curve shows the post-Orientale $R$ plot.
The green curve is the $R$ plot for the post-mare Copernican and Eratosthenian craters; the green straight line is a best-fit power law.  The orange curve shows a simulated lunar highlands crater density curve reduced by 99\%.  The
 ``Combination'' curve (black) is the combination of
post-mare Copernican and Etatosthenian craters and 1\%  the lunar highlands; the black straight line is a  best-fit power law. 
}
\label{fig:fig16}
\end{figure}

Some recent studies argue that the LHB may have extended to much more recent times than 3.8 or 3.7 Ga in the inner Solar system.  An analysis of impact generated spherules in pre-Cambrian terrestrial sediments indicates that large basin-forming impacts continued on Earth much longer than previously thought, possibly up to $\sim 2$ Ga \citep{Johnson:2012}.  \cite{Bottke:2012} have suggested, based on a dynamical model of the orbital migration history of the giant planets, that the LHB began 4.1 Ga and continued to produce basin-size craters upto at least $\sim$~2.5 Ga;  
they suggest that most of the late impactors could have originated in an extended and now largely extinct portion of the asteroid belt between 1.7 and 2.1 AU, a so-called E-belt.  
From the impact flux estimates of \cite{Bottke:2012}'s model (ten lunar basins between 3.7~Ga and 4.1~Ga, fifteen terrestrial basins between 2.5~Ga and 3.7~Ga), and adopting their 17:1 ratio of the gravitational cross section of Earth and Moon, it is straightforward to calculate the ratio of the modeled average impactor flux during the LHB epoch (4.1--3.7 Ga) to that during the extended LHB epoch (2.5--3.7 Ga):
\begin{equation}
{\hbox{average impactor flux (LHB)}\over\hbox{average impactor flux (extended LHB)}} = 
{17\times10/(4.1-3.7)\over 15/(3.7-2.5)} =34.
\end{equation}
This implies that the extended LHB impact flux was very significantly smaller than the peak LHB flux.
The crater record provides additional useful constraints for this extended-LHB model, as we discuss below.

We know from the post-Orientale crater counts that the LHB was still occurring after the Orientale impact, but at a rate about 6.5 times less than the peak period (see Figure~\ref{fig:fig05}).
Based on crater counts and radiometric ages of lunar samples, the age of the lunar maria is from 3.9 to 1.2
Ga with the greatest lava eruption volume occurring between about 3.3~Ga and 3.7~Ga
\citep{Hiesinger:2000,Hiesinger:2003}. The superposed craters on the maria
have a size distribution consistent with Population 2 (see Figure \ref{fig:fig04}).
Therefore, the impact rate of the LHB extension must have been low enough that the accumulation of
Population 2 masked the later stages of Population 1 impacts. Since
Population 1 is deficient in smaller craters ($<50$ km diameter)
compared to Population 2 (see Figures~\ref{fig:fig01} and \ref{fig:fig02}),
this would result in much less modification of Population 2 at smaller
diameters. Furthermore, the formation of parts of the lunar maria
during the time interval 3.9 Ga to 2.0 Ga would have destroyed some of
the later Population 1 craters.   As a simple illustration, we show in Figure \ref{fig:fig16}
the results of a simulation of the
effects of a 99{\%} reduction of the impact flux of the LHB between 3.9~Ga
and 2.0 Ga and its combination with the post mare Population 2
craters. The resulting crater size distribution is similar to the post-mare
Population 2 curve in both shape and magnitude of crater density. 
The power law fits to the curves are also similar in slope and magnitude, but the  ``Combination” curve is at a slightly higher density and slightly greater slope than the post-mare curve. This indicates that the LHB impact flux was reduced by $\sim99\%$ or more between 3.9 Ga and 2.0 Ga. 

Another recent study, \cite{Morbidelli:2012}, employs cosmogonic models of the ancient asteroidal population and its dynamical evolution to argue that the LHB began as an uptick of a factor of 5-to-10 in the bombardment rate at 4.1 Ga and decayed with an exponential timescale of $\sim144$ myr. This differs only slightly from the timeline of the `terminal lunar cataclysm' of an impact flux spike during 4.0-3.8 Ga inferred by \cite{Tera:1974} and others based on radiometric analyses of lunar samples.  The crater SFDs presented here do not conclusively distinguish between the two timelines.  We can only note that the crater record indicates that the transition from Population 1 to Population 2 projectiles was quite complete by about 3.7 Ga because Population 1 is absent on the lunar maria that formed since that time.

The uncertainty of the onset age and the duration of the LHB do not affect our main result of two populations of impactors and their association with the LHB and the post-LHB bombardment in the inner solar system. 

\subsection{Dynamical Mechanism for the LHB}
\label{subsec:LHBdynamics}

The congruence of the size distribution of the projectiles of
Population 1 craters and of the MBAs indicates a dynamical ejection process
that was largely insensitive to the asteroid mass, and very distinct
from the dynamical mechanism that produced Population 2.
\citet{Strom:2005} suggested that Population 1 could be identified with the
LHB, and that the source of the LHB impactors was the main asteroid
belt.  The LHB impact spike could plausibly have been caused by the
size-independent dynamical ejection of main belt asteroids during a
short-duration orbital migration of the giant planets.

The orbital migration of the giant planets was previously proposed to
explain the orbit of Pluto and to predict the orbital distribution in
the Kuiper Belt \citep{Malhotra:1993,Malhotra:1995} and to explain the
relative paucity of asteroids in the outer asteroid belt
\citep{Liou:1997}. The hypothesis of giant planet migration has
subsequently been supported with discoveries in the Kuiper Belt as
well as subsequent theoretical studies \citep[cf.][]{Morbidelli:2009a}.
With regard to the main asteroid belt, it has long been noted
that there exist many `gaps', known as the `Kirkwood Gaps'
\citep{Kirkwood:1882}, near the locations of many mean motion resonances
with Jupiter and the $\nu_6$ secular resonance associated with Saturn's
mean perihelion precession rate.  The dynamical effect of these
gravitational resonances is that they cause orbital instabilities over
certain limited ranges of semimajor axis in the main asteroid belt.
The orbital migration of Jupiter and Saturn would have caused these
unstable zones to sweep across a range of asteroid semimajor axes that
were previously populated with asteroids, thereby causing asteroids to
be ejected from the main belt into planet-crossing orbits.  Indeed,
the de-biased orbital distribution of the main asteroid belt reveals
that the extent of the Kirkwood gaps and the density of asteroids near
the $\nu_6$ secular resonance cannot be explained with the perturbations of
the giant planets in their current orbits, but can be accounted for
only if Jupiter has migrated inward by $\sim 0.2$ AU and Saturn has migrated
outward by $\sim 1$ AU \citep{Minton:2009}.  Furthermore, from the
eccentricity distribution of main belt asteroids, it is inferred that
the timescale of Jupiter and Saturn's migration was possibly as short
as a few million years \citep{Minton:2011}.  If planet
migration is the correct explanation, then one also needs to explain
its short timescale as well as the nearly 600 million year delay
between the formation of the giant planets and their orbital
migration.  Such an explanation has been proposed by
\citet{Tsiganis:2005} with a scenario known as the ``Nice model''.
In this scenario, the giant planets initially form in a marginally
stable orbital configuration and migrate very slowly for the first few
hundred million years, until such time as Jupiter and Saturn encounter
a 2:1 mean motion resonance.  This planet-planet resonant encounter
causes a strong chaotic episode in the orbital evolution of all the
giant planets, changing their orbital eccentricities and triggering a
fast migration.  This causes a major dynamical instability in both the
Kuiper Belt and the asteroid belt, and, therefore, a spike in the
impact flux of both comets and asteroids on the inner Solar system
planets and the Moon \citep{Gomes:2005}. The authors estimate that
the cataclysmic bombardment lasted 30--150 million years, and that
comet impacts dominated at early times and asteroid impacts dominated
at later times during the impact spike.   
Additional work is needed to fully test this model and to constrain its free parameters \citep[e.g.][]{Dawson:2012,Agnor:2012}.

An alternative interpretation is offered by \citet{Cuk:2010,Cuk:2011,Cuk:2012}.
These authors argue that the transition from
Population 1 impactors to Population 2 impactors occurred prior to the
formation of the Imbrium and Orientale basins, i.e., prior to $\sim$~3.8 Ga,
and therefore the LHB must be associated with Population 2, rather
than Population 1 craters. \citet{Cuk:2012} presents the following scenario.
The Population 1 craters were made over an extended period of time
prior to $\sim 3.8$ Ga, and the source of their impactors was a very large
primordial population of Mars-crossing asteroids 
that decayed gradually over several hundred million years. The
LHB was caused by the singular break-up at $\sim$~3.8 Ga of a Vesta-size
body in this population.  The size distribution of the break-up
fragments is postulated to be similar to that of Population 2.  Some
basic aspects of this scenario are consistent with the crater record
as we understand it: the existence of two different impactor
populations, and the similarity of the ancient Population 1 with the
size distribution of the main belt asteroids (very plausibly the
putative primordial Mars-crossing asteroid population as well as the 
E-belt shared the main belt size distribution).
But some aspects directly contradict the data as we understand it.
Population 2 craters do not dominate the Imbrian and post-Orientale craters
(see Figure \ref{fig:fig05}, also \citet{Malhotra:2011}).
Second, there is a conflict between Population 2 being that
of a large-asteroid-break-up event causing a short-lived LHB and the
evidence that the Population 2 SFD has been in
near-steady-state over the past $\sim$~3.8 gigayears of the post-LHB crater
record. Third, the expected fragment size distribution of asteroid
break-up events, based on observations of asteroid collisional
families and on numerical simulations of family-formation, is quite
different than the SFD of either Population 1 or
Population 2 projectiles \citep[e.g.][]{Benavidez:2012}.
A large-asteroid-break-up event as a
dynamical cause of the LHB has also been investigated previously by
\citet{Zappala:1998} and \citet{Ito-Malhotra:2006}; the latter work
 concluded that this was not a viable mechanism
because it requires an implausibly large asteroid parent body.
For these reasons, this scenario is not supported by the data
as we understand it.
  
\subsection{Implications for Age Dating from the Impact Crater Record}
\label{subsec:implication}

If the Late Heavy Bombardment is the result of a cataclysmic event, as
the evidence indicates, then the previous cratering record has been
significantly obliterated, and the ancient impact flux (prior to about 4~Ga) is presently
unknown. Therefore, our current knowledge of the impact flux history
in the inner Solar system from the impact crater record is not
adequate to date surfaces older than about 3.9 billion years. Studies
that claim to date surfaces older than this date from the cratering
record of Population 1 are only dating them between about 3.8 and 4.0
billion years.

Surfaces that display Population 2 craters are younger than about 3.7
billion years and can be dated relatively reliably by using the NEO flux at the planet in question \citep[e.g.][]{Lefeuvre:2011}.
There may be impact craters formed during the
extended LHB as discussed earlier in which case the derived model age
will be an upper limit. Although comet impacts are surely contained
within the Population 2 crater population, they have not been abundant
enough to affect the SFD. Therefore, Population 2 must
be dominated by asteroid impacts, unless comet impacts produce the
same crater SFD as NEOs. However, ages derived from the
NEO flux are upper limits because some comet impacts are probably present.

The small-crater population (approximately $D<1$ km diameter on the Moon
and Mars, $D > 10$ km on Mercury) should be used with great caution to
date surfaces because it is contaminated by large numbers of secondary
impact craters \citep{McEwen:2006,Robbins:2011,Xiao-Strom:2012}.
This is particularly true for Mercury where the
secondaries are larger for a given size crater than anywhere else in
the Solar system \citep{Xiao:2014}. Some craters on Mercury have
more circular secondaries, rendering the distinguishing of primaries
and secondaries more difficult than on the other planets \citep{Xiao:2014}.
Mercury basin secondaries begin to affect the crater SFD
at diameters of about 9--10 km in almost all heavily cratered areas of
the planet \citep{Strom:2011}.

\section{Summary}
\label{sec:summary}
\begin{figure}[ht] \centering
\includegraphics[width=\myfigwidthL, angle=0]{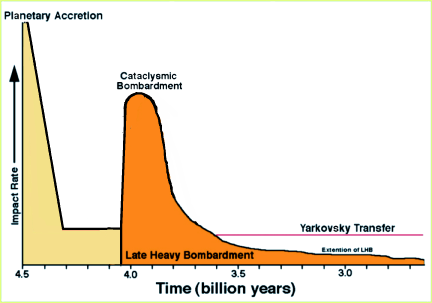}
\caption[]{%
Schematic diagram of the impact history of the inner Solar system.
The high impact rate near 4.5 Ga represents planetary accretion that
formed the terrestrial planets. The steep drop in the impact rate just
after 4.5 Ga represents the fast collisional and dynamical loss of the
local impactor population after planet formation.
The LHB began at some time prior to about 3.9 Ga,
 as indicated by the evidence presented in
sections \protect{\ref{sec:terrestrial}} and \protect{\ref{sec:discussion}}.
The red horizontal lines between about 4.4 and 4.1 billion years
and after $\sim 3.6$ billion years represent impacts by means of Yarkovsky
transfer of asteroids from the asteroid belt; the former is higher
than after the LHB because there were many more asteroids at that time.
The total number of impacts via Yarkovsky transfer after the
main period of the LHB would have masked the putative extended LHB
(see section~\ref{subsec:LHBage} and Figure \protect{\ref{fig:fig16}}).
}
\label{fig:fig17}
\end{figure}

Two populations of objects of distinctly different size-frequency distributions 
have impacted the inner Solar system planets and the Moon. 
When combined with the accumulated data on the age-dating of lunar 
and meteorite samples, as well as insights from Solar system dynamics, 
the simplest interpretation is the following. 
One population is responsible for the Late Heavy Bombardment and the other is
responsible for impacts after the Late Heavy Bombardment.
The population responsible for the Late Heavy Bombardment originated from
Main Belt Asteroids while the younger population originated from Near
Earth Objects. That the size distribution of the projectiles
responsible for the Late Heavy Bombardment is the same as Main Belt
Asteroids means that they were ejected in a size-independent manner by
means of a gravitational instability. A plausible cause was the
orbital migration of Jupiter and Saturn causing a sweeping of
gravitational resonances through the Main Asteroid Belt and resulting
in a cataclysmic bombardment of the inner Solar system. The younger
population is also derived from the Main Asteroid Belt, but ejected by
the size-dependant Yarkovsky effect that gradually feeds
asteroids into unstable gravitational resonances; we observe the
source of these impactors at the present time as the NEOs.
Figure \ref{fig:fig17} is a diagram to illustrate in a very general way
the impact history of the inner Solar system. 
Surfaces younger than about 3.7 Ga can be
dated in a relatively reliable way by measuring crater densities and
using estimates of the near-planet asteroid impact flux at the appropriate
planet. But this technique must be applied in crater diameter ranges larger
than those of Population S (secondaries). These ages will be upper
limits because some comet impacts and extended LHB impacts are
possibly present in Population 2 craters.  The ancient crater record
prior to the LHB has been significantly obliterated, and ancient
surfaces cannot be reliably age-dated from the cratering record.

\normalem
\begin{acknowledgements}
We thank Z. Ivezic and J. S. Stuart for providing us with digital
versions of their published data.  RM acknowledges research support from NSF grant \#AST-1312498.
We also thank the anonymous referee for comments which improved the quality of this paper.

\end{acknowledgements}

\clearpage
\bibliographystyle{raa}
\bibliography{renu}

\includepdf[pages=1-2]{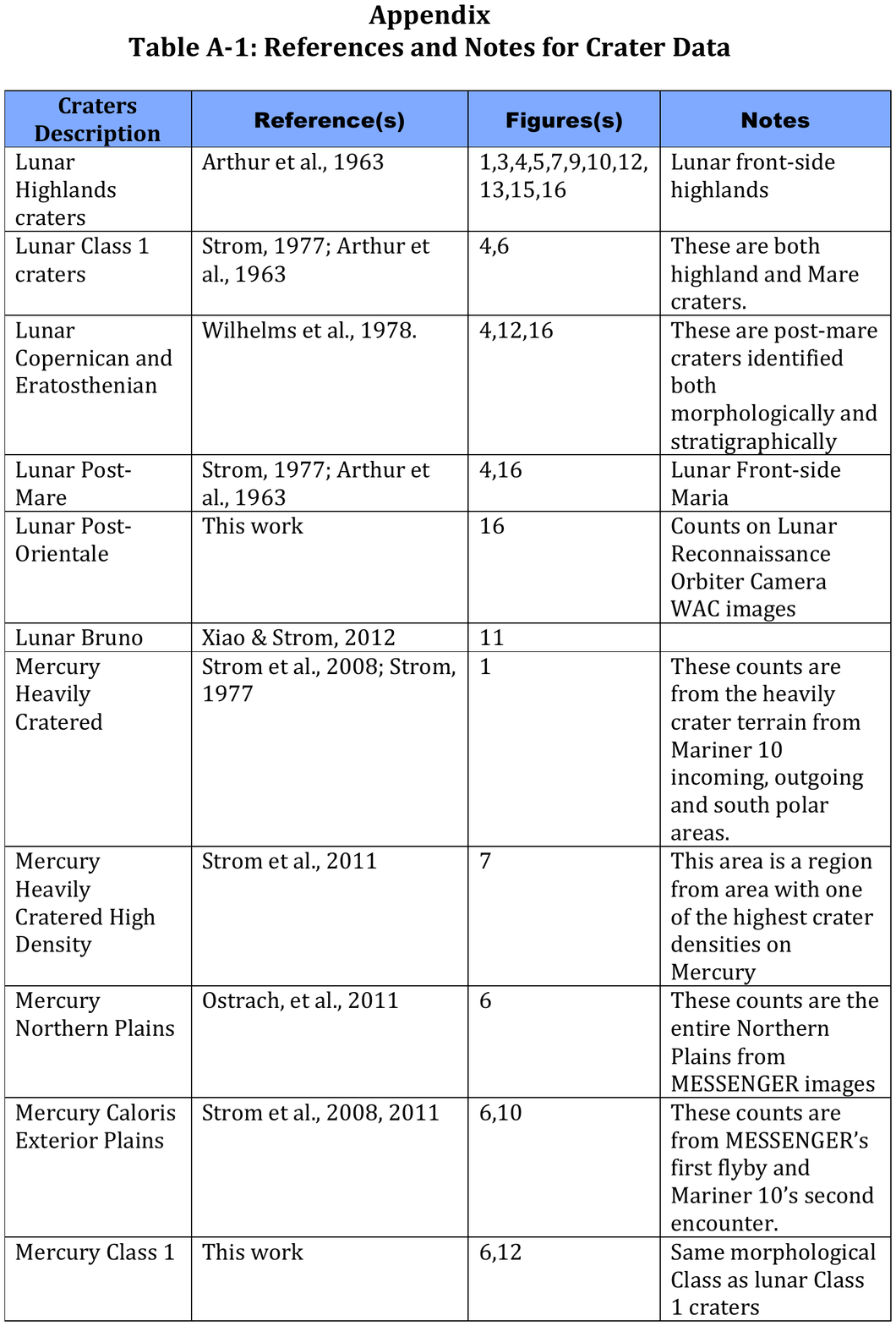}
\includepdf{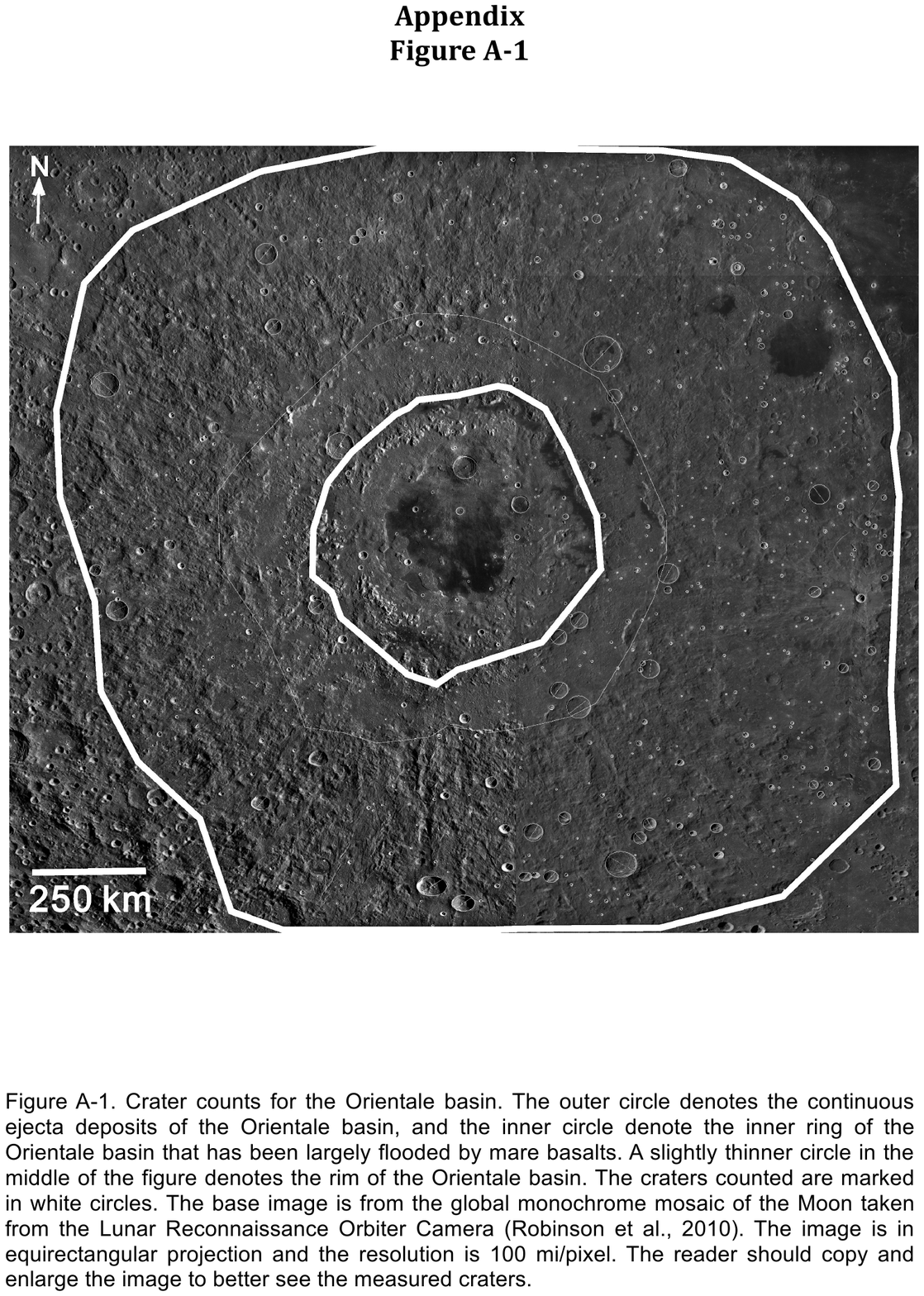}
\includepdf{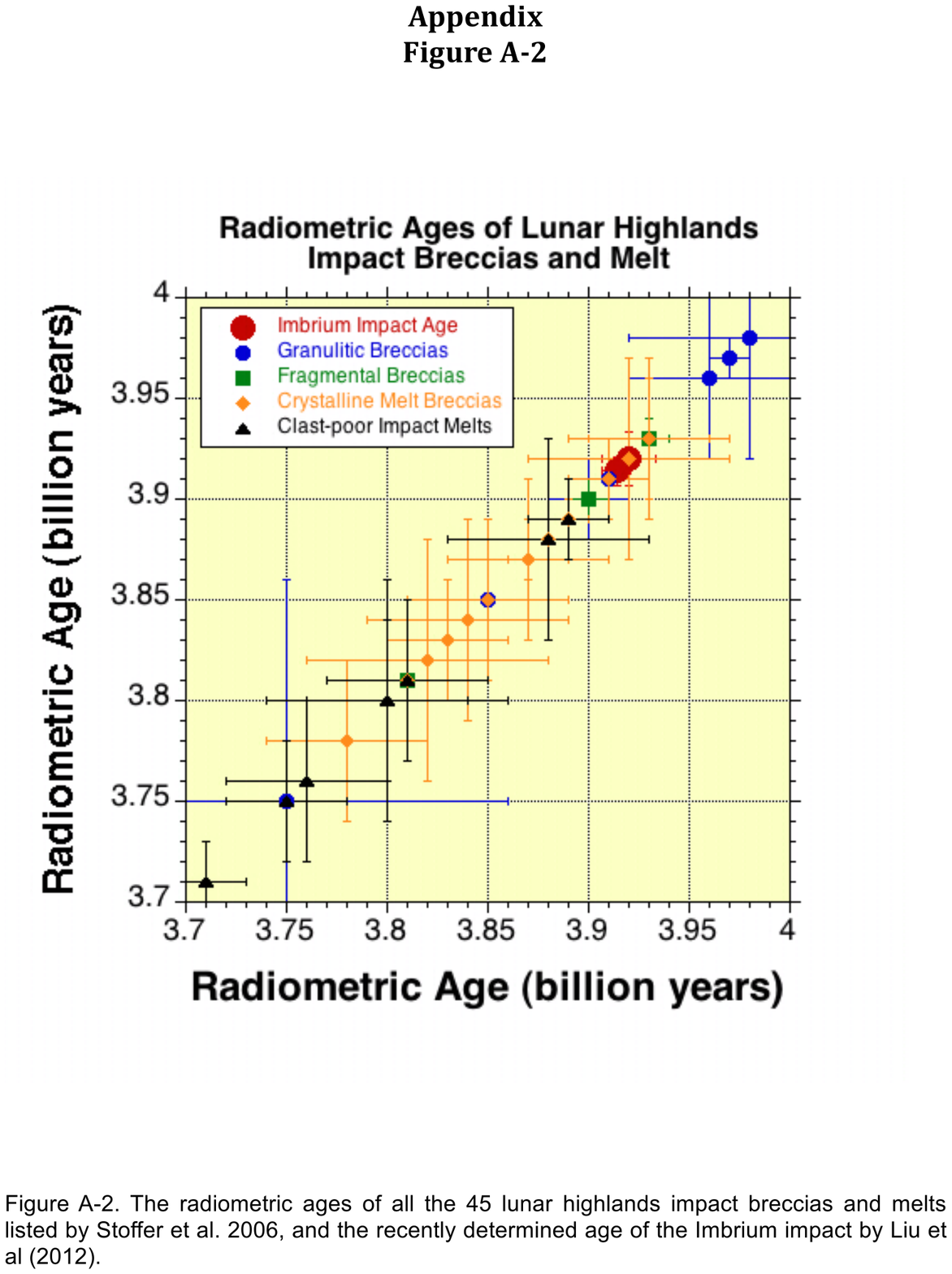}
\end{document}